\documentclass[
,final            
  ]
  {aipproc}

\layoutstyle{6x9}

\usepackage{amsfonts,amssymb,amsmath,latexsym,epsfig,euscript}

\newcommand{\pd}{\partial}

\DeclareMathOperator{\sign}{sign}

\newcommand{\Dd}{\EuScript{D}^2}
\newcommand{\Dh}{\Dd_\mathrm{H}}

\newcommand{\DD}[1]{\Dd_{\mathrm{H}_{#1}}}

 \newcommand{\ap}{\alpha^{\prime}}

\begin{document}

\title{
\vskip -50pt
{\begin{normalsize}
\mbox{} \hspace*{100mm}\hfill DAMTP-2007-90 \\
\vskip 30pt
\end{normalsize}}
Rolling Tachyon in Nonlocal Cosmology
}



\author{L. Joukovskaya}{
  address={DAMTP, Centre for Mathematical Sciences, University of Cambridge,
Wilberforce Road, Cambridge CB3 0WA, UK}
}

\begin{abstract}
Nonlocal cosmological models derived from String Field Theory are considered.
A new method for constructing rolling tachyon solutions in the FRW metric
in two field configuration is
proposed and solutions of the Friedman equations with nonlocal
operator are presented. The cosmological properties of these
solutions are discussed.
\end{abstract}

\maketitle

\section{Introduction}

In recent works there appears interest in nonlocal cosmological
models derived from String Field Theory in connection with
problem of describing cosmological inflation or accelerating
expansion of the Universe.

Modern cosmological data indicates that expansion of the
Universe is accelerating. It may be owing to a component of the
Universe with negative pressure, Dark Energy. Recent results of
WMAP \cite{cosmo-obser} together with Ia supernovae data give
us the following range for the dark energy state parameter
$w=-0.97^{+0.07}_{-0.09}$. From theoretical point of view
it is natural to distinguish three essential different cases, $w>-1$,
$w=-1$ and $w<-1$. The third one is more challenging one because
it violates all natural energy conditions and there might appear problems
of an instability at classical and quantum level.
Recently it was proposed  to consider $D$-brane decay in cubic superstring
field theory as a model of cosmological Dark Energy \cite{IA_marion}.
At the same time there have been a number of attempts to
realize description of the early Universe via nonlocal
cosmological models (for more details see \cite{LJ} and refs therein).

In this work we will continue research \cite{IA_marion,LJ,AJ-JHEP,AJV-JHEP}
of properties of nonlocal cosmological models derived from string field
theory. More precisely, we will consider a cosmological model
based on gravity interacting with tachyon matter governed by
the tachyon action in CSSFT when fields up to zero mass are taken into account \cite{AJK}.

\section{Setup}

To describe open string states living on a single non-BPS
$\mathrm{D}$-brane one has to consider GSO$\pm$ states \cite{Sen}.
GSO$-$ states are Grassmann even, while GSO$+$ states are
Grassmann odd.
The low level action \cite{AJK} contains the tachyon field $\phi$ and one
auxiliary field $u$
\begin{equation}
S[u,\phi]=\frac{1}{g_o^2\alpha^{\,\prime \frac{(p+1)}{2}}} \int\;
d^{p+1}x\left[ u^2(x)- \frac{\ap}2
\partial _\mu \phi(x)\partial ^\mu \phi(x)+\frac14 \phi^2(x) -
\frac{1}{3\gamma^2} U(x)\Phi^2(x)
\right], \label{action-x}
\end{equation}
where the $\Phi$ and $U$ are defined as
$\Phi=e^{k \Box} \phi$, $U=e^{k \Box} u$, $k=-\ap \ln \gamma$,
$\gamma=\frac{4}{3\sqrt{3}}$.

Covariant generalization of action (\ref{action-x}) after rescaling of fields and
coordinates in dimensionless space-time variables has the form
\begin{equation}
S=\int d^4x\sqrt{-g}\left(\frac{m_p^2}{2}R+
\frac{\xi^2}{2}\psi \square_g \psi
+\frac{1}{2}\psi^2+\frac{1}{4}\upsilon^2-\frac{1}{2}\Upsilon \Psi^2-\Lambda^\prime-T \right),
\end{equation}
where $g$ is the metric,
$\square_g=\frac1{\sqrt{-g}}\pd_{\mu}\sqrt{-g}g^{\mu\nu}\pd_{\nu}$,
$\psi$ is a dimensionless scalar tachyon field,
$\Psi=e^{\frac{1}{8} \square_g}\phi$, $\upsilon$ is a dimensionless auxiliary
scalar field, $\Upsilon=e^{\frac{1}{8} \square_g}\upsilon $,
$m_p^2=g_4 \frac{M_p^2}{M_s^2}$, $M_p$ is a Planck mass,
$M_s$ is a characteristic string scale, $g_4$ is a dimensionless four
dimensional effective coupling constant, $\Lambda^\prime $ is an
effective  cosmological constant, $T$ is the
brane tension.
In this work we will be interested in the case of the FRW metric
$
ds^2={}-dt^2+a^2(t)\left(dx_1^2+dx_2^2+dx_3^2\right).
$
We will consider spatially homogeneous configurations for which
Beltrami-Laplace operator takes the form
$\Box_g=-\partial_t^2-3H(t)\partial_t$. For the convenience of numerical
calculations let us introduce the following notation $\Dh=\partial_t^2+3H(t)\partial_t$.
The corresponding  equations of motion have the form
\begin{subequations}
\label{two-fields}
\begin{equation}
e^{\frac{1}{4} \Dh} \Upsilon(t)=\Psi^2(t),
\end{equation}
\begin{equation}
\left(-\xi^2\Dh+1\right)e^{\frac{1}{4} \Dh}\Psi(t) = \Upsilon
\Psi (t),
\end{equation}
\end{subequations}
and corresponding Friedman equations have the following form
\begin{subequations}
\label{fr}
\begin{equation}
\label{fr-1}
3H^2=\frac{1}{m_p^2}~(\frac{\xi^2}{2} (\partial \psi)^2- \frac{1}{4}\upsilon^2
-\frac12 \psi^2+  \frac{1}{ 2}\Upsilon \Psi^2+E_{nl_1}+E_{nl_2}+E_{nl_3}+E_{nl_4}+
\Lambda^\prime+T),
\end{equation}
\begin{equation}
\label{dotH}
\dot H={}-\frac{1}{m_p^2}~(\frac{\xi^2}{2} (\partial \psi)^2+E_{nl_2}+E_{nl_4}),
\end{equation}
\end{subequations}
where
\begin{subequations}
\label{4nl}
\begin{eqnarray}
\label{nl1} E_{nl1}(t)&=& \frac18\int_0^1 d\rho\left(
e^{\frac{2-\rho}{8}\Dh}(-\xi^2 \Dh+1) \Psi\right)
  \left(\Dh  e^{\frac{\rho}{8}\Dh} \Psi
\right),\\
\label{nl2} E_{nl2}(t)&=&- \frac18\int_0^1 d\rho\left(
\partial e^{\frac{2-\rho}{8}\Dh}(-\xi^2 \Dh+1)\Psi\right)
  \left(\partial e^{\frac{\rho}{8} \Dh}\Psi
\right),\\
\label{nl3} E_{nl3}(t)&=& \frac{1}{16} \int_0^1 d\rho\left(
e^{\frac{2-\rho}{8} \Dh}\Upsilon \right) \left(\Dh
e^{\frac{\rho}{8}\Dh}\Upsilon\right),
\\
\label{nl4} E_{nl4}(t)&=& - \frac{1}{16} \int_0^1 d\rho\left(
\partial e^{\frac{2-\rho}{8} \Dh} \Upsilon
\right)\left(\partial e^{\frac{\rho}{8}\Dh} \Upsilon\right).
\end{eqnarray}
\end{subequations}

\section{Rolling Tachyon in the FRW Universe}

\begin{figure}
\centering
\includegraphics[width=39mm]{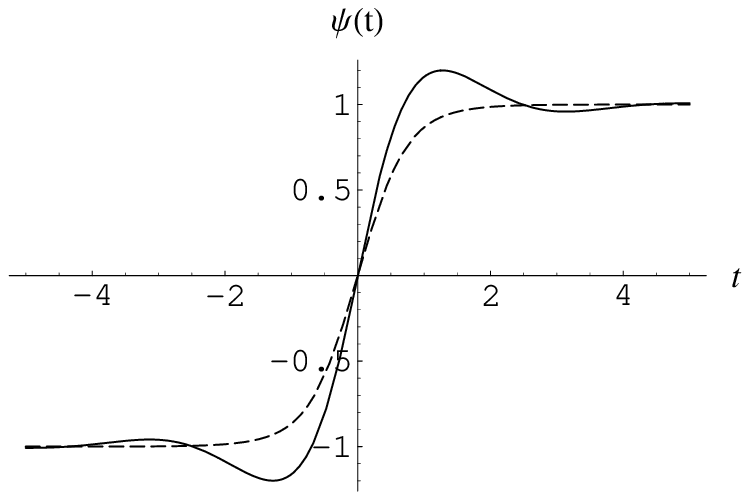}~~~~~~~~
\includegraphics[width=39mm]{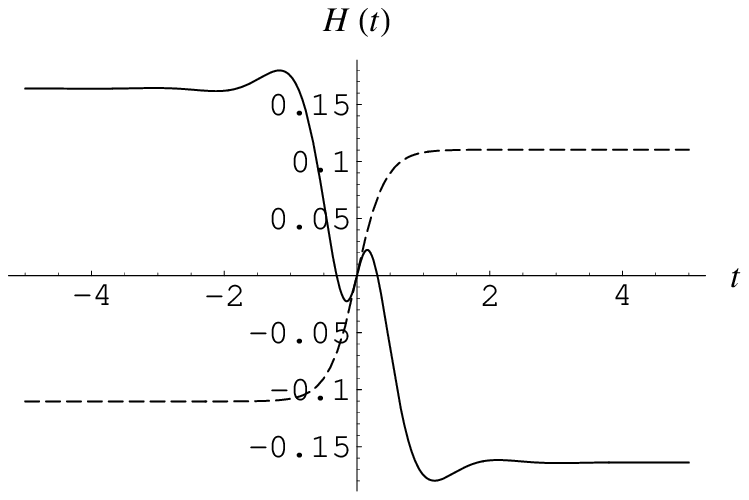}~~~~~~~~
\includegraphics[width=39mm]{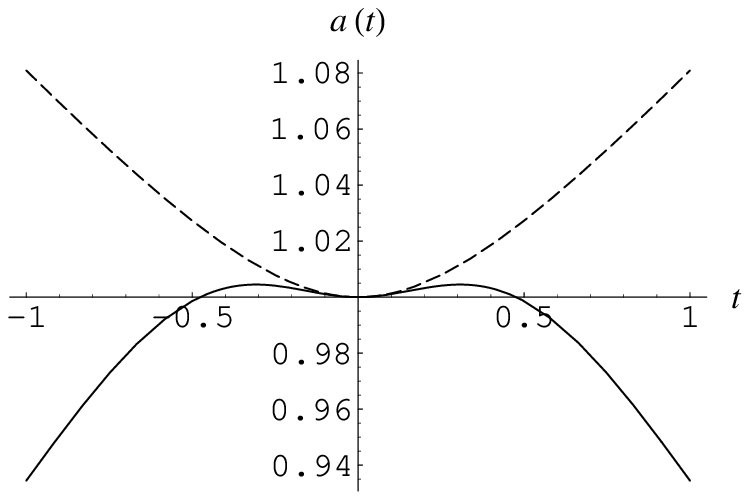}
\caption{Solutions of the Friedmann equations $\psi$,
$H$ and $a$  for approximation $u \approx e^{k \Box} u$ with $\xi^2=0$ (dashed line) and
$\xi^2=-\frac{1}{4\ln\gamma}$ (solid line).}
\label{fig:one-field}
\end{figure}

The approximation $u \approx e^{k \Box} u$ in the action (\ref{action-x})
for Minkowski case was considered in \cite{IA_marion,LJ},  it simplifies scalar field equation and
with $\xi^2=0$ reproduces $p$-adic model for $p=3$ which is also interesting for
the applications. Recently for example it was proposed to consider $p$-adic
theory as a theory of inflation \cite{BarnabyBiswasCline}.
This approximation was studied in Minkowski space in \cite{AJK,Yar-JPA}
and recently in FRW space in \cite{LJ}. In this approximation the auxiliary field $u$
(or $\upsilon$ in  terms of rescaled fields) can be integrated and we will
have only one equation of motion  $\left(-\xi^2\Dh+1\right)e^{\frac{1}{4}\Dh}\Psi =\Psi^3$,
simplification of Friedmann equations also takes place (see \cite{LJ} for details).
Figure \ref{fig:one-field} presents solutions
$\psi$, $H$ and $a$ for $\xi^2=0$ and $\xi^2=-\frac{1}{4\ln\gamma}$.

\begin{figure}
\centering
\includegraphics[width=39mm]{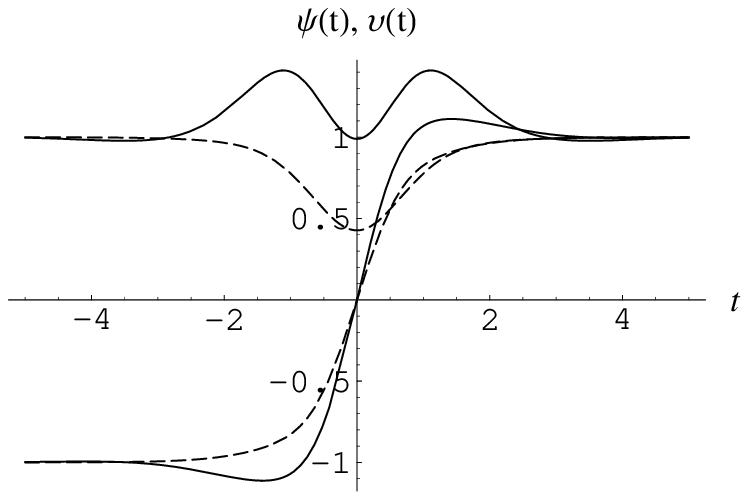}~~~~~~~~
\includegraphics[width=39mm]{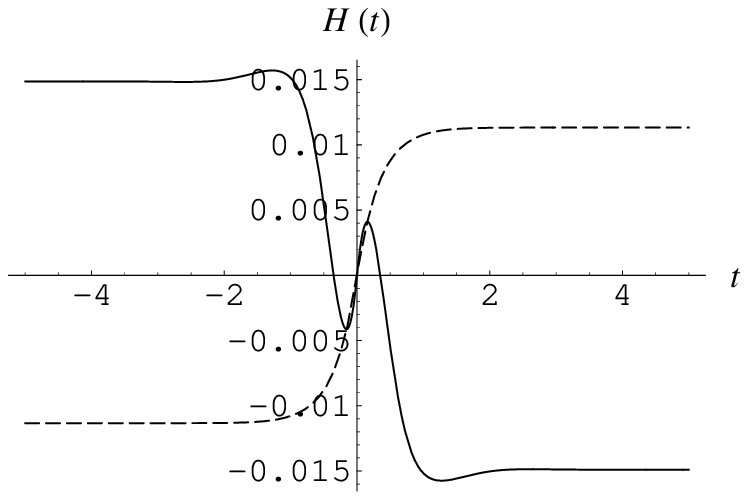}~~~~~~~~
\includegraphics[width=39mm]{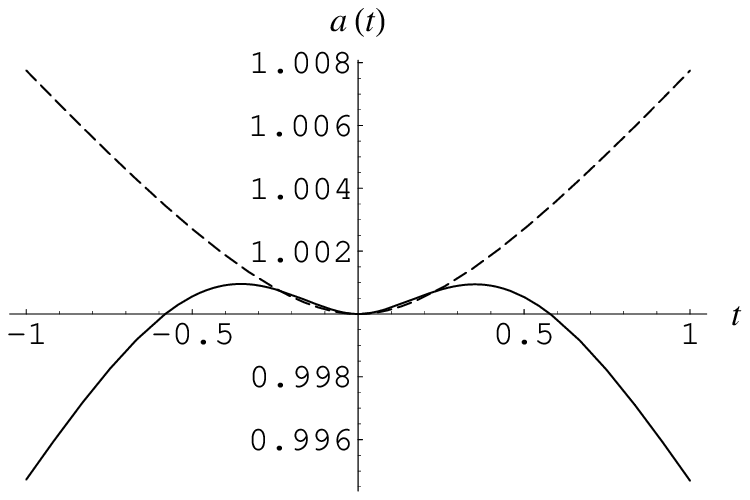}
\caption{Solutions of the Friedmann  equations (\ref{two-fields}), (\ref{fr})
with $\xi^2=0$ (dashed line) and
$\xi^2=-\frac{1}{4\ln\gamma}$ (solid line).
On the most left  figure $\psi$ has odd kink type shape,
while $\upsilon$ has even wavelet type shape.}
\label{fig:two-fileds}
\end{figure}

Let us return to the case of two field configuration \cite{AJK,two-fields}
$\Psi$ and $\Upsilon$ which is more interesting.
To find solutions of the system of equations
(\ref{two-fields}) and integrating equation (\ref{dotH}),
$H(t)=-\frac{1}{m_p^2} \int_0^t d \tau ~\dot H(\tau)$, we construct the following iterative process
$$
\Psi = \lim_{n\to\infty} \Psi_n, ~~\Upsilon = \lim_{n\to\infty} \Upsilon_n,
~~H = \lim_{n\to\infty} H_n,
$$
where iterations $\Psi_n$, $\Upsilon_n$, and $H_n$ are in turn obtained as limits
of sub-iterations
$
\Psi_{n+1} = \lim_{m\to\infty} \Psi_{n,m},
~~\Upsilon_{n+1} = \lim_{m\to\infty} \Upsilon_{n,m}, ~~H_{n+1} = \lim_{m\to\infty} H_{n,m},
$
which are recursively defined as ($m \geqslant 0$)
$$
\Upsilon_{n,m+1}=\frac{(-\xi^2\DD{n}+1)e^{ \frac{1}{4} \DD{n}}\Psi_{n,m}}{\Psi_{n,m}},~~~~~
\Psi_{n,m+1}=\sign(t)\sqrt{e^{ \frac{1}{4} \DD{n}}\Upsilon_{n,m+1}},
$$
$$
H_{n,m+1}=-\frac{1}{m_p^2}\int_0^t d\tau
\left[
  \frac{\xi^2}{2}(\pd e^{\frac{1}{8}\DD{n,m}}\Psi_{n+1})^2
-\frac{1}{16} \int_0^1 d\rho\left(
\partial e^{\frac{2-\rho}{8} \DD{n,m}} \Upsilon_{n+1}
\right)\times
\right.
$$
$$
\left.
\left(\partial e^{\frac{\rho}{8}\DD{n,m}} \Upsilon_{n+1}\right)-
\frac18\int_0^1 d\rho\left(
\partial (-\xi^2 \DD{n,m}+1)e^{\frac{2-\rho}{8}\DD{n,m}} \Psi_{n+1}\right)
\left(\partial e^{\frac{\rho}{8} \DD{n,m}}\Psi_{n+1}
\right)
\right].
$$
The initial iterations in $m$ is taken as
$
\Phi_{n,0}=\Phi_n, ~~H_{n,0}=H_n;
$
$
\Psi_0(t)=\tanh(t),~~H_0(t)=0
$ (see \cite{LJ} for the details of analogous iterative process).
Obtained solutions are presented on the Fig. \ref{fig:two-fileds}.
For $\xi^2=0$ we have nonsingular accelerating Universe with a bounce. Hubble
function goes to the constant as $t \to \pm \infty$, hence the state parameter
tends to $w=-1$ as $t \to \pm \infty$. Behavior of the Hubble function
for $\xi^2=0.96$ also illustrates the possibility to obtain
the state parameter which is close to $-1$.

Note that  $\Lambda^\prime$ does not enter equations (\ref{two-fields}), (\ref{dotH}),
but can be determined from (\ref{fr-1}). Following Sen's conjecture we put $D$-brane's tension to
$T=-V(\Psi=\pm1,\Upsilon=1)=\frac{1}{4}$) \cite{IA_marion, AJK} and
thus $\Lambda^\prime$ is determined uniquely for each field
configuration in the same way as in \cite{LJ}.

In this work we have studied the properties of nonlocal cosmological models
derived  from String Field Theory in the Friedmann space-time. We have construct
new method for solving partial differential equations with infinitely many time
derivatives, which constitute a new class of mathematical physics equations and are
recently discussed in the literature \cite{Yar-JPA,BarnabyBiswasCline,mathphys}.
We obtained classical solutions of the corresponding Friedmann equations which
can be considered as a first approximation to the quantum solutions and might be useful
for the study of ways to avoid the cosmological singularity problem \cite{Ekpyrotic}.

\begin{theacknowledgments}
The author is grateful to Neil Turok for his support.
The author also would like to thank I.~Aref'eva, G.~Calcagni, D.~Mulryne, Ya.~Volovich
for  fruitful discussions.
The author gratefully acknowledge the use of the UK National
Supercomputer, COSMOS, funded by PPARC, HEFCE and Silicon
Graphics. This work is supported by  the Center for Theoretical
Cosmology, in Cambridge and in part by RFBR grant 05-01-00758,
INTAS grant 03-51-6346  and Russian President's grant
NSh--672.2006.1.
\end{theacknowledgments}

\end{document}